\documentclass[prd,aps,floats,twocolumn]{revtex4}
\usepackage{epsfig}
\usepackage{color}

\newcommand{\be}{\begin{equation}}
\newcommand{\ee}{\end{equation}}
\newcommand{\ba}{\begin{eqnarray}}
\newcommand{\ea}{\end{eqnarray}}
\newcommand{\bc}{}

\newcommand{\bea}{\begin{eqnarray}}
\newcommand{\eea}{\end{eqnarray}}
\newcommand{\cf}{F_{K}}
\newcommand{\ccf}{F_{KK}}

\begin{document}

\preprint{\
\begin{tabular}{rr}
\end{tabular}
}
\title{Vector field models of modified gravity and the dark sector}
\author{J.~Zuntz$^{1}$, T.G~Zlosnik$^{2}$, F.~Bourliot$^{3}$,
P.G.~Ferreira$^{1}$, G.D.~Starkman${^4}$ }
%
\affiliation{
$^1$Astrophysics, University of Oxford, Denys Wilkinson Building, Keble Road, Oxford OX1 3RH, UK\\
$^2$Perimeter Institute for Theoretical Physics, 31 Caroline Street North, Waterloo, Ontario N2L 2Y5, Canada\\
$^3$ CPHT, Ecole Polytechnique, 91128 Palaiseau cedex, France \\
$^4$Department of Physics/CERCA/ISO,Case Western Reserve University, Cleveland,OH, 44106-7079
}

\begin{abstract}
We present a comprehensive investigation of cosmological constraints on the class of vector field formulations of modified gravity called Generalized Einstein-Aether models.  Using linear perturbation theory we generate cosmic microwave background and large-scale structure spectra for general parameters of the theory, and then constrain them in various ways.  We investigate two parameter regimes: a dark-matter candidate where the vector field sources structure formation, and a dark-energy candidate where it causes late-time acceleration.  We find that the dark matter candidate does not fit the data, and identify five physical problems that can restrict this and other theories of dark matter.  The dark energy candidate does fit the data, and we constrain its fundamental parameters; most notably we find that the theory's kinetic index parameter $n_{\mathrm{ae}}$  can differ significantly from its $\Lambda$CDM value.

\end{abstract}

\date{\today}
\pacs{PACS Numbers : }
\maketitle

\section{Introduction}

Over the past few years, a new suite of models for the dark sector has been proposed. They
invoke a vector field which is normally constrained to lie along the time-like direction and
may lead to modifications to the gravitational sector \cite{ZlosnikFerreiraStarkman2007,Maroto2008, DimopoulosEtAl2006,Koivisto:2008xf}. Sometimes called
Einstein-Aether models, they tend to entangle two of the main paradigms currently being
considered: on the one hand modified gravity and on the other dark matter and energy.

Vector field models are attractive because they seem to be able to resolve the problem of
the dark sector (i.e. dark matter {\it and} energy) in a unified way. Most of the emphasis 
has gone into constraining vector field models that lead to accelerated expansion
\cite{Maroto2008, DimopoulosEtAl2006, Koivisto:2008xf} although there is a fair amount of work for which the
the vector field leads to a relativistic version of Modified Newtonian Dynamics (MOND)
\cite{Milgrom1983} or can even play the role of dark matter \cite{ZlosnikFerreiraStarkman2007,Zhao2007}. 
In fact vector field models seem to incorporate what seems to be a generic feature 
of relativistic modified gravity models \cite{Science2009}: that it is impossible to construct
relativistic models that just modify the gravitational sector without introducing new degrees of freedom, which can then behave like either dark matter or dark energy (although for other approaches see, for example, \cite{Milgrom:2009gv,Blanchet:2009zu,Blanchet:2008fj,Soussa:2003sc,Bruneton:2008fk,Sagi:2009kd}).

There has been significant progress in trying to constrain these models. For example, at a fundamental level it
has been shown that a broad class will lead to instabilities and the formation of caustics, 
signaling a break down of the fundamental theory \cite{Withers2008}. It has also been shown
that for a general choice of kinetic terms, these theories will be plagued by ghosts or tachyons
\cite{Carroll2009, Eling:2005zq, Lim:2004js}. These pathologies are worrying but do not entirely rule out vector field models-
it has been shown that modifications to the kinetic term, for example, can cure them.

Substantial work has been done on understanding how these fields in these theories behave on macroscopic
scales, either through their interaction with matter to form galaxies and clusters \cite{Chang2008}, or on the largest scales, affecting the growth of structure and its
effect on the CMB \cite{ZlosnikFerreiraStarkman2008}. Indeed for a particular, ``vanilla'' version of the
vector-field model, detailed and definitive constraints have been placed on the
various coupling constants \cite{ZuntzFerreiraZlosnik2008, Jacobson:2008aj}. 

In \cite{ZlosnikFerreiraStarkman2008}, it was found that one of the key effects that vectors
would have would be to modify the growth rate of structure. This is not surprising- theories
that modify gravity tend to have this effect. We also found that it lead to a mismatch
between the two gravitational potentials a potentially observable effect \cite{ZhangBean2007}. In
this paper we wish to pursue this analysis and quantify how strong these effects are. Although
we focus on a particular (albeit broad) class of theories, we are interested in extracting
{\it general} lessons from these models. We believe that much of what we learn by looking at these models will shed light on other models of modified gravity (such as, for example, $f(R)$ theories
\cite{Bean2007} and bimetric theories \cite{Bekenstein2004a}).

The structure of this paper is as follows. In Section \ref{theory} we lay out the essential ingredients for a reasonably broad class of vector-like models and its background evolution, and specialize to the form used in the remainder of the paper.  In section \ref{linear} we lay out the equations of the perturbed theory, and how we implement them with theoretical constraints.  In section \ref{as dark matter} we discuss the problems with modelling dark matter with the theory.  In section \ref{as dark energy} we find and constrain the parameters which let the theory act as  dark energy.  In section \ref{conclusions} we conclude and draw more general lessons about the dark sector.

\section{The Theory}
\label{theory}
\subsection{Theory Definition}

A general action for a vector field $A^{a}$ coupled to gravity can
be written in the form:
\begin{eqnarray}
S=\int d^4x \sqrt{-g}\left[\frac{R}{16\pi G}+{\cal L}(g^{ab},A^{b})\right]
+S_{M} \label{genaction}
\end{eqnarray}
where $g_{ab}$ is the metric, $R$ the Ricci scalar of that metric,
$S_M$ the matter action and $\cal{L}$ is constructed to be generally
covariant and local. By construction $S_M$ only couples to the metric, $g_{ab}$ and 
{\it not}
to $A^{a}$. 

We will restrict ourselves to consider a Lagrangian that
only depends on covariant derivatives of $A$ and we will consider a unit time-like $A^{a}$. Such a theory can be written
in the form:
\begin{eqnarray}
\label{eq:Lagrangian}
{\cal L}(g^{ab},A^{a})&=&\frac{M^2}{16\pi G}
	 F(K) +\frac{1}{16\pi G}\lambda(A^a A_a+1)
	 \\
K&=&M^{-2}K^{ab}_{\phantom{ab}mn}
\nabla_a A^{m}\nabla_b A^{n}   \\
\label{tensor}
K^{ab}_{\phantom{ab}mn}&=&c_1g^{ab}g_{mn}
+c_2\delta^a_{\phantom{a}m}\delta^b_{\phantom{b}n}+
c_3 \delta^a_{\phantom{a}n}\delta^b_{\phantom{b}m}
\end{eqnarray}
where $c_i$ are dimensionless constants and $M$
has the dimension of mass.  We have removed an additional $c_4$ `acceleration' term in accordance with the transformation described in \cite{ElingJacobson2006}.

As was the case with TeVeS, the constant $G$ may be a different number from 
the locally measured value of Newton's gravitational constant.  $\lambda$ is a non-dynamical 
Lagrange-multiplier
field with dimensions of mass-squared.

The gravitational field equations for this theory, obtained by varying the action with respect to $g^{ab}$ (see \cite{ZlosnikFerreiraStarkman2007}) are
\begin{equation} \label{Einsteineq1}
G_{ab}=\tilde{T}_{ab}+8\pi GT^{\mathrm{matter}}_{ab}
\label{fieldI}
\end{equation}
where the stress-energy tensor for the vector field is given by
\begin{eqnarray}
\tilde{T}_{ab} &=& \frac{1}{2}\nabla_{c}
(F_{K}(J_{(a}^{\phantom{a}c}A_{b)}-
J^{c}_{\phantom{c}(a}A_{b)}-J_{(ab)}A^{c}))
\nonumber \label{lagrangian equation} \\ && -F_{K}Y_{(ab)}
+\frac{1}{2}g_{ab}M^{2}F+\lambda A_{a}A_{b}
 \\
F_{K} &\equiv & \frac{dF}{dK}\\
J^{a}_{\phantom{a}c} &=&
(K^{ab}_{\phantom{ab}cd}+
K^{ba}_{\phantom{ba}dc})\nabla_{b}A^{d}
\end{eqnarray}
Brackets around indices denote
symmetrization\footnote[2]{we adopt the convention $X_{(ab)}=\frac{1}{2}(X_{ab}+X_{ba})$, $X_{[ab)}=\frac{1}{2}(X_{ab}-X_{ba})$} and $Y_{ab}$ is the functional derivative
\begin{eqnarray}
Y_{ab} =\nabla_{c}A^{e}\nabla_{d}A^{f}
\frac{\delta(K^{cd}_{\phantom{cd}ef})}{\delta 
g^{ab}}
\end{eqnarray}

The equations of motion for the vector field, obtained by varying with respect to $A^{b}$ are
\begin{eqnarray}
\label{eq:veceq1}
\nabla_{a}(F_{K}J^{a}_{\phantom{a}b})
+F_{K} y_{b}&=&2\lambda A_{b}
\label{vectoreom}
\end{eqnarray}
where once again we define the functional derivative
\begin{eqnarray}
y_{b}=\nabla_{c}A^{e}\nabla_{d}A^{f}
\frac{\delta(K^{cd}_{\phantom{cd}ef})}
{\delta A^b}
\end{eqnarray}

Finally, variations of the action with respect to $\lambda$ will fix $A^b A_b=-1$. By inspection, contracting both sides of (\ref{eq:veceq1}) 
with $A^{b}$ leads to a solution for $\lambda$ in terms of the the vector field
and its covariant derivatives.

These equations allow us to study a general theory of the form presented
in equation \ref{genaction} with a unit time-like vector field. 
For our particular, restricted choice
of $K$ we have $Y_{ab}=-c_{1}\left[ (\nabla_{c}A_{a}(\nabla^{c}A_{b})-(\nabla_{a}A_{c})(\nabla_{b}A^{c})\right]$ and $y_b=0$.

\subsection{Background Cosmology}
\label{background}

In this paper we will restrict ourselves to background cosmologies where the spacetime is of the spatially flat Friedmann-Robertson-Walker (henceforth FRW) form:

\be
\label{lel}
g_{ab}dx^{a}dx^{b} =-{dt}^{2}+a(t)^{2}\delta_{ij}dx^{i}dx^{j} .
\ee

The energy momentum content of each matter fluid $(i)$ in the background is taken to be of the usual form

\begin{eqnarray}
T^{i}_{ab}=(P^{i}+\rho^{i})u^i_{a}u^i_{b}+P^{i}g_{ab}.
\end{eqnarray}

Where $\rho^{i}$ and $P^{i}$ are the co-moving density and pressure of the $i$th fluid respectively.
We assume that all fluids have identical four velocities $u^{a}=(1,0,0,0)$. Furthermore we define $\rho \equiv \sum_{i}\rho^{(i)}$.

In spacetimes with FRW symmetries, the vector field must align with the direction $\partial_{t}$ and so the vector field is entirely fixed to have components $(1,0,0,0)$ in the co-ordinate system (\ref{lel}).  Explicitly, the background value of the scalar $K$ is given by:
\begin{equation}
K_{\mathrm{FRW}}= 3\frac{\alpha H^{2}}{M^{2}}
\label{KFRW equation}
\end{equation}

Where $H\equiv \partial_{t}\ln(a(t))$ and:
\begin{equation}
\alpha \equiv c_{1}+3c_{2}+c_{3}.
\label{alpha definition}
\end{equation}

The Friedmann equation then takes the form \cite{ZlosnikFerreiraStarkman2008}: 

\begin{eqnarray}
\label{fried}
\left[1-\alpha K^{1/2}\frac{d}{d K}\left(\frac F{ K^{1/2}}\right)\right]H^2 &=& \frac{8\pi G}{3}\rho \label{00dyn}
\end{eqnarray}
where $\rho$ still includes only the matter components.

We note that the combination of (\ref{00dyn}) and the aether stress energy tensor allows us to write down an effective energy density $\rho_{\mathrm{ae}}$ and pressure $P_{\mathrm{ae}}$ of the aether. From this we may define the fractional energy density  $\Omega_{\mathrm{ae}}\equiv 8\pi G\rho_{\mathrm{ae}}/3H_{0}^{2}$, and the aether equation of state parameter $w_{\mathrm{ae}}\equiv P_{\mathrm{ae}}/\rho_{\mathrm{ae}}$:

\begin{eqnarray}
\Omega_{\mathrm{ae}} &=& \frac{M^{2}}{6}\left[\frac{\partial}{\partial H}\left(\frac{F}{H}\right)\right]_{H=H_{0}}
\end{eqnarray}

\begin{equation}
w_{\mathrm{ae}} = -1-\frac{1}{3H^{2}}\frac{\frac{d^{2}}{dtdH}F}{\frac{d}{dH}
\left(\frac{F}{H}\right)}
\end{equation}

\begin{figure}
\begin{center}
\epsfig{file=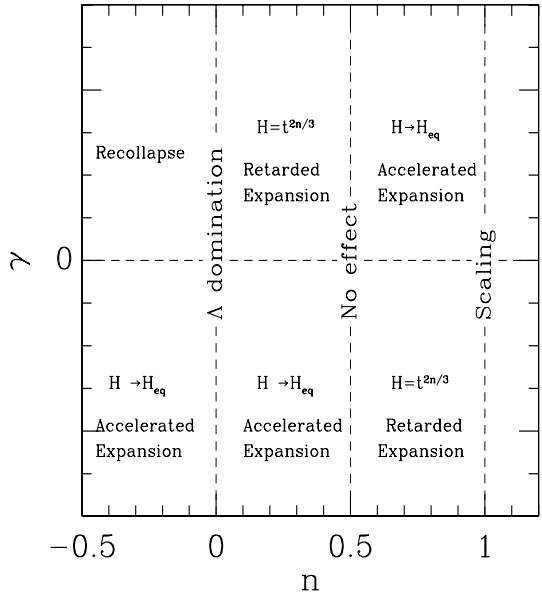,width=8.3cm,height=9.3cm}
\caption{A schematic representation of the types of the late-time background evolution permitted by the functional form $F=\gamma (-K)^{n}$ as a function of $(n,\gamma)$ for $n<1$.}
\label{fig1dyn}
\end{center}
\end{figure}

The effect of the vector field on the background expansion may be see
as an expansion rate dependent modification to Newton's constant  i.e. schematically (\ref{fried}) is an equation of the form $3H^{2}=8\pi G(H^{2})\rho$. It was found in \cite{ZlosnikFerreiraStarkman2008} that various forms of the function $F$ permitted a wide variety of cosmological dynamics:
the presence of the vector field variously leading to accelerated expansion, slowed expansion, rescaling of $G$, and recollapse as summarized in Figure \ref{fig1dyn}

\subsection{The functional form of $F(K)$ in cosmology}
\label{F}

We must now specify the form of the function $F(K)$ in equation (\ref{lagrangian equation}).  There is an obvious set of candidates here - we could attempt to be directly consistent with MOND and use the same branch of $F$ as it uses on small scales.  In appendix \ref{mond regime bad} we show that doing so would make it impossible to consistently generate late-time acceleration behavior in the background cosmology.  Instead, we will use a simple and reasonable ansatz that works for the regime $|K|\gg1$ that we consider here.

Existing functional forms for F(K) in the MOND regime typically are dominated for a single monomial term for $|K|\gg1$ (see for instance \cite{Famaey:2005fd}) and so it seems reasonable to restrict the function to take this form: 
\begin{eqnarray}
F = \gamma (-K)^{n_{\mathrm{ae}}} \  \ &K&<0\nonumber \\
F =  \gamma (+K)^{n_{\mathrm{ae}}}  \  \ &K& >0 
\label{kinetic equation}
\end{eqnarray}
where ${n_{\mathrm{ae}}}\leq 1$.  This form has sufficient power to express a wide variety of behavior, and the parameters $\gamma$ and $n$ shall be central to our further analysis.

\section{Linear perturbation Theory}
\subsection{Formalism \& Theory}
\label{linear}

We have seen that the vector field can have a significant effect on the quasistatic, weak field limit and the background cosmological geometry.
Significant evidence for non-baryonic mass persists on the largest cosmological scales \cite{DodelsonLiguori2006} therefore it is vital that a relativistic theory of MOND can account for this. As mentioned, it has been argued \cite{DaiMatsuoStarkman2008b} that even in the quasistatic, weak field limit, a spatial tilt to the vector field may produce significant deviations under some circumstances from the local MOND force law. 
Similarly it was shown \cite{ZlosnikFerreiraStarkman2008} that in the context of linear cosmological perturbations the energy density associated with the projection of the vector field onto surfaces of constant conformal time could, to a degree, act as a cosmological `dark matter' .
In this paper we shall comprehensively address the question of whether the Lagrangian (\ref{eq:Lagrangian}) represents a viable model of the dark sector in light of precision cosmology.

In the main body of the paper we will consider scalar perturbations. In Appendix \ref{vt} we derive the equations of motion for the vector field's two divergenceless vector modes as well as the gravitational wave tensor modes (the speed of propagation of which is in general modified by the vector field). Requiring the stability of these modes puts constraints on (\ref{eq:Lagrangian}) (see Section \ref{cons}).

We shall work in the synchronous gauge (see for instance \cite{Ma:1995ey}) and so the metric takes the following form:

\begin{equation}
g_{\mu\nu}dx^{\mu}dx^{\nu}=-a^{2}d\tau^{2}+a^{2}[\gamma_{ij}+h_{ij}]dx^{i}dx^{j}
\end{equation}

where $\tau$ is conformal time, $\gamma_{ij}$ is a spatially flat spacelike 3-metric perturbed by $h_{ij}$ which is built from two scalar potentials  $\eta$ and $h$:

\begin{equation}
h_{ij}(\textbf{x},\tau)= \int d^{3}k e^{i\textbf{k}\cdot\textbf{x}}[\hat{\textbf{k}}_{i}\hat{\textbf{k}}_{j}h(\textbf{k},\tau)+(\hat{\textbf{k}}_{i}\hat{\textbf{k}}_{j}-\frac{1}{3}\delta_{ij})6\eta(\textbf{k},\tau)]
\end{equation}

Similarly we will expand the aether field as:

\begin{equation}
A^{\mu}=\frac{1}{a}(1,\partial_{i}V)
\end{equation}

The zeroth component of the aether field is, by virtue of the gauge choice and the constraint,
fixed as equal to $1$ up to second order in perturbations.

A slight complication in the field equations arises because of the presence of the function $F$ which depends nonlinearly on the scalar $K$. We assume that for modes of interest one may consistently regard the perturbation to $K$ as being much less than unity. Thus one can expand $K$ as $K = K^{0}+ K^{\epsilon}$ and  $F$ as $F= F^{0}+F_{K}^{0}K^{\epsilon}$ where $K^{\epsilon} \ll 1$ where the superscript $0$ denotes the quantity corresponding to a function's background value. Explicitly we have that:

\begin{eqnarray}
K^{0} &=& 3\frac{\alpha {\cal H}^{2}}{a^{2}M^{2}}\\
{\cal H}K^{\epsilon} &=& -\frac{2}{3}K^{0}(k^{2}V-\frac{h'}{2}) 
\end{eqnarray}

Where primes denote derivatives with respect to conformal time and we have used the conformal Hubble parameter ${\cal H}\equiv H/a$.
We also make use of the following identity:

\begin{eqnarray}
(K^{0})'{\cal H}=-4K^{0}({\cal H}^{2}-\frac{1}{2}\frac{a''}{a})
\end{eqnarray}

Henceforth we will drop the superscripts on  $K^{0}$  and $F^{0}$ i.e. $K$ and $F$ shall be assumed to represent \emph{background} values of the fields.

Towards simplifying the form of the equations, we will will rather use the field $\xi$ instead of $V$, where $\xi$ is defined as:

\begin{equation}
\xi \equiv V-\frac{1}{2k^{2}}(h+6\eta)'
\end{equation}

For further compactness of notation we define the variables:

\begin{eqnarray}
\hat{\alpha} &\equiv &\left(1+2\frac{F_{KK}}{F_{K}}K\right)\alpha \\
\hat{c}_{1} &\equiv & \left(1+2\frac{F_{KK}}{F_{K}}K\right)c_{1}
\end{eqnarray}

The vector field equation of motion (\ref{eq:veceq1}) becomes :

\begin{eqnarray}
0 &=& c_{1}(1+c_{13}F_{K})\frac{(F_{K}\xi')'}{F_{K}}+2{\cal H}c_{1}(1+c_{13}F_{K})\frac{(F_{K}\xi)'}{F_{K}} \nonumber \\
\nonumber &&+ [2c_{1}(1+c_{13}F_{K})(\frac{a''}{a}-{\cal H}^{2})+2(\hat{c}_{1}+\hat{\alpha})({\cal H}^{2}-\frac{1}{2}\frac{a''}{a})\\
\nonumber &&+c_{1}c_{13}(F_{KK}K')'+\frac{1}{3}(\hat{\alpha}+2c_{13})k^{2}]\xi \\
\nonumber && +(c_{1}+\hat{\alpha})\eta'+(\hat{c}_{1}+\hat{\alpha})\frac{1}{k^{2}}({\cal H}^{2}-\frac{1}{2}\frac{a''}{a})(h'+6\eta')\\
 && -\frac{3}{2}\frac{c_{1}}{k^{2}}\frac{(F_{K}\Sigma_{f})'}{F_{K}} \label{eq:veceq}
\end{eqnarray}

The relevant Einstein equations are:

\begin{eqnarray}
\label{eq:ein1}
(1-\frac{1}{2}\hat{\alpha} F_{K})k^{2}\eta' &=& 4\pi G a^{2}
                       i k^{j}\delta T^{0}_{\phantom{0}j} \nonumber \\
&&                        +\frac{1}{6}k^{4}(\hat{\alpha}+2c_{13})F_{K}\xi 
\end{eqnarray}

and

\begin{eqnarray}
\label{eq:ein2}
(1+\frac{1}{2}c_{1}F_{K})({\cal H}h'-2k^{2}\eta) &=& -8\pi Ga^{2}\delta T^{0}_{\phantom{0}0} \nonumber\\
\nonumber && -\frac{1}{2}F_{K}(c_{1}+\hat{\alpha})6{\cal H}\eta'\\
\nonumber && -2\alpha F_{KK}K{\cal H}k^{2}\xi\\
\nonumber && +\frac{F_{K}c_{1}k^{2}}{a^{2}}\left( a^{2}(1+c_{13}F_{K})\xi \right)'\\
&& -\frac{3}{2}c_{1}F_{K}\Sigma_{f}
\end{eqnarray}

Where we have used the fact that:

\begin{eqnarray}
\label{shear}
(h+6\eta)''+2{\cal H}(h+6\eta)'-2k^{2}\eta &=& -3\Sigma_{f} \nonumber \\
&&+2c_{13}k^{2}[F_{K}(2{\cal H}\xi+\xi')\nonumber \\
&& +F_{KK}K'\xi]
\end{eqnarray}

The functions $\delta T^{0}_{\phantom{0}j}$ and $\delta T^{0}_{\phantom{0}0}$ are the first order perturbations to the corresponding components of the matter fields' stress energy tensors. Summation over field species is assumed. The field $\Sigma_{f}$ is the scalar component of the total fluid shear i.e. $\Sigma_{f}=-8\pi G a^{2}(\hat{k}_{i}\hat{k}^{j}-\frac{1}{3}\delta^{j}_{\phantom{j}i})\Sigma^{i}_{\phantom{i}j}$ and $\Sigma^{i}_{\phantom{i}j}\equiv \delta T^{i}_{\phantom{i}j}-\frac{1}{3}\delta^{i}_{\phantom{i}j}\delta T^{k}_{\phantom{k}k}$. A gauge invariant formulation of the theory's equations may be found in \cite{LiMotaBarrow2007}. 

\subsection{Parameter Constraints}
\label{cons}

We can immediately see a number of constraints on the $c_{i}$ and the form of the function $F$. From (\ref{eq:veceq}) it can be shown that for in the limit of timescales shorter than a Hubble time the quantity:
\begin{equation}
C_{S}^{2} = \frac{2}{3}\frac{(\frac{\hat{\alpha}}{2}+c_{13})}{c_{1}(1+c_{13}F_{K})} 
\end{equation}
can be interpreted as the squared sound speed of the field $\xi$. The avoidance of exponentially growing subhorizon modes dictates that $C_{S}^{2}$ should be positive definite. 

Similarly, one may consider the field equations of the two divergenceless `vector' modes of the vector field and the two transverse traceless `tensor' modes of of the metric. Each respectively has a squared sound speed function (named $C_{V}^{2}$ and $C_{T}^{2}$ respectively) which, as in the scalar case, should be positive definite. These functions are calculated in Appendix \ref{vt} and are as follows:

\be
C_T^2=\frac{1}{1+c_{13}F_{K}}
\ee

\be
C_V^2=\frac{\cf}{2c_1}\frac{2c_1+\cf(c_1^2-c_3^2)}{1+c_{13}F_{K}}
\ee

Note that the gravitational wave tensor modes now generically have a time dependent speed of propagation.

Collectively the three positivity constraints imply the following constraints on the $c_{i}$ and function $F$ parameter space:

\begin{eqnarray}
1+\cf c_{13} & > & 0\\
(\hat{\alpha}+2c_{13})/2c_{1} & > & 0 \\
\cf(1+F_{K}\frac{c_1^2-c_3^2}{2c_1})& > & 0
\end{eqnarray}

Now we turn to the Einstein equations (\ref{eq:ein1}) and (\ref{eq:ein2}). Terms in the perturbed vector field stress energy tensor $\delta \tilde{T}_{ab}$ may contain terms proportional (up to a time-dependent function of the background fields) to $\delta G_{ab}$ and indeed this can be seen in (\ref{eq:ein1}) and (\ref{eq:ein2}) through the appearance of terms proportional to the $c_{i}$ on the left hand sides of the equations.
If the effect $\xi$ is negligible then the effect of the aether stress energy terms proportional to components of the Einstein tensor may always be absorbed into a redefinition of Newton's constant as a time-dependent effective gravitational coupling. Thus, even if $\xi$ has comparatively little effect, there may be a considerable modification to the link between the matter fields and the gravitational field.
The resulting gravitational couplings should be greater than or equal to zero otherwise the gravitational field will interpret normal matter as violating energy conditions, and so risking the appearance of instabilities. This restriction implies the following constraints:

\begin{eqnarray}
(1-\frac{1}{2}\hat{\alpha}F_{K})> 0
\end{eqnarray}

\begin{eqnarray}
(1+\frac{1}{2}c_{1}F_{K}) > 0
\end{eqnarray}

Throughout our analysis, we will only consider regions of the model's parameter space which satisfy these constraints.
\subsection{Computation}
\label{computation}
To study the effects of the vector field in detail we have modified the structure formation Boltzmann code {\sc CMBEASY} \cite{Doran:2003sy}.  We add a Newton-Raphson solver for the Hubble parameter, with added aether components.  The perturbation evolution is also modified to include the aether components $\xi$ and $\xi'$, and their contribution to the density, pressure and shear perturbations.  We also include the altered metric perturbations in the calculation of the CMB source function.  We use adiabatic initial conditions\cite{ZlosnikAetherModes_preparation}. Since Boltzmann codes are very highly optimized for $\Lambda$CDM models, care must be taken to ensure that modifications are performed in a consistent manner - for example an unmodified Friedmann equation is often assumed for computational efficiency.

To explore parameter spaces, {\sc CMBEASY} is coupled to a Monte-Carlo Markov Chain (MCMC) engine \cite{Doran:2003ua}.  We extended this engine to include our new Aether parameters: $\Omega_{\mathrm{ae}}$, $c_+=(c_1+c_3)$, $c_-=(c_1-c_3)$, $C_S^2$, $n_{\mathrm{ae}}$ and  $M$.  We include the full ranges of these parameters by allowing the kinetic term to take two branches, for positive and negative $K$ as in (\ref{kinetic equation}). 

We constrain models using both CMB data from the WMAP experiment \cite{Komatsu:2008hk} and large-scale structure from the SDSS survey \cite{SDSS:2007wu}, though not always at the same time.  As we shall see, some Aether models are extremely poor fits to the combined data sets; to illustrate such problems we want to find models that fit only the large scale structure data.  In regimes where fits are extremely poor, MCMC does not work particularly well.  We ameliorate such situations by running larger numbers of shorter Markov chains and re-starting from their best-fit positions, and sometimes by abandoning MCMC altogether and simply performing random searches for good parameter combinations.

\section{The vector field as Dark Matter}
\label{as dark matter}
\subsection{Background evolution \& doppler peak positions}
\label{background subsection}
An unusual property of the model considered here is that at the level of cosmological perturbations the field can mimic a perturbed pressureless fluid in the formation of large scale structure whilst behaving entirely differently in the cosmological background  \cite{ZlosnikFerreiraStarkman2008} . In this paper we would like to consider not just large scale structure but also other cosmological probes. The anisotropies in the CMB temperature are sensitive to the background dynamics and perturbed dynamics of source of cosmic mass discrepancies in a largely distinguishable manner. 

If the aether plays the role of dark matter only in the perturbations, then the background expansion is dark energy-dominated at an earlier time.  For a given $H_0$ this reduces the expansion rate of the universe between recombination and now and so decreases the angle subtended on the sky by given distance at last scattering.  This moves the CMB doppler peaks to higher $\ell$.  Although the aether can give a suitable time dependence to the effective gravitational coupling $G$ in the Friedmann equations so as to yield the same expansion rate as $\Lambda$CDM,  the necessary functional forms of $F$ are extremely contrived. For instance these forms essentially contain a new constant scale roughly equal to the Hubble parameter at matter-radiation equality.  Even in the event of such a construction, it may not be possible for the squared sound speed of vector field perturbations $C_{S}^{2}$ to remain sufficiently small as to behave like cold dark matter in perturbations \cite{ZlosnikFerreiraStarkman2007}.

The other input to the peak position is the sound horizon at last scattering.  The physics of this is sufficiently robust that changes compensatory to the alteration in the distance to last scattering are not feasible without exceptional fine tuning of the perturbational behavior of the aether.  Indeed, it has been argued that only additional components that behave like non-relativistic matter in the background might fix this problem \cite{Ferreira:2008ma}. 

Could there indeed be such a non-relativistic matter in the background which allows acoustic peak positions which are consistent with the data whilst leaving the aether in its role as the seed of the formation of large scale structure? A appropriate candidate would appear to be massive neutrinos. A suitable mass of such particles so as to account for the right effective contribution to the dust component of the background typically implies that the neutrinos themselves are unable to clump on small enough scales so as to be a good candidate for all of the dark matter. 
Such a solution as has been proposed in \cite{SkordisEtAl2006} and \cite{Angus2008b}. 

\subsection{Perturbation evolution}
We now argue that even with the inclusion of massive neutrinos, one generically expects the aether to have an unacceptable influence on the large scale CMB anisotropy if it is to also play the dominant role in structure formation.  

The first requirement for successful perturbation evolution is that structure can form at all.  One necessary condition for this is that the sound speed of the structure seed not be too large, since this would wash out structure.  We require that the sound horizon in the model be less than the smallest scales where structure can form linearly: $C_S k_{\mathrm{max}} \tau  \lesssim 1$, where $k_\mathrm{max} \sim 0.2 h/M\mathrm{pc}$.  For matter power observations at $\tau \sim 3\times 10^{4}$, the present epoch, this yields $C_S \lesssim 10^{-4}$. 

There are two underlying physical processes that further constrain the models.

The first is a change to the growth rate of perturbation amplitude.  This can cause discrepancies between the amplitudes we expect in the matter power spectrum and the CMB, since the evolution between the two is different.  It can also lead to an integrated Sachs-Wolfe (ISW - see below) effect during the matter era since $\Phi$ will accrue a time dependence.

The second is the increased presence of a $\Phi-\Psi$ metric shear.  This also leads (directly) to a matter era ISW.

\subsection*{Observable 1: ISW}
\label{ISW subsection}
Under the assumption of adiabaticity we have that
the anisotropy in the CMB, $\Delta T({\hat n})/T$ on large scales in a given spatial direction 
${\hat n}$ is given by

\begin{equation}
\label{isw equation}
\frac{\Delta T({\hat n})}{T}\simeq -\frac{1}{3}\Psi(\tau_*,d_*{\hat n}) -
\int_{\tau_*}^{\tau_0}d\tilde\tau({\Psi'}+{\Phi'})[\tilde\tau,(\tau_0-\tilde\tau){\hat n}]
\end{equation}
where $\tau_*$ is the conformal time of last scattering, $\tau_{0}$ is the conformal time today, $d_*$ is the comoving radius of the surface of last scattering, and $\Phi$ and $\Psi$ are the conformal Newtonian gauge gravitational potentials.  The integral in (\ref{isw equation}) is the integrated Sachs-Wolfe (ISW) effect.  Writing the integrand of (\ref{isw equation}) as $\Psi' + \Phi' = -(\Phi-\Psi)' + 2\Phi'$, we can see this as time derivatives of a shear part and a growth rate part.

In the standard cosmological model,  the field $\Phi$ has negligible time dependence during matter domination.  It gains a time dependence only when the background starts accelerating, and only then can the resulting growth rate ISW contribution be considerable. In the model considered here, the situation may be rather different.  It was found in \cite{ZlosnikFerreiraStarkman2008} that substantial contributions to the ISW may occur even during the matter era.  For the aether field to seed structure formation the field $\xi$ must have a suitable growing mode solution in the matter era.  Typically the corresponding spatial curvature perturbation $k^{2}\Phi$ will then have a time dependence via the Poisson equation. 

The shear part can also gain a time dependence in the aether model, which in the $\Lambda$CDM is very small even during acceleration.

Each of these effects depend on the functional form of $F$, the time-dependence of the $\xi$ growing mode and the choice of the parameters $c_{i}$. It is extremely challenging to find combinations of the parameters which allow for a realistic growth of structure whilst making the ISW acceptably low. This is most easily illustrated by considering the theory TeVeS \cite{Bekenstein2004a} which has many of the same properties as the model considered here. It may be shown that TeVeS can be written as a single metric theory with a timelike vector field of unfixed norm \cite{ZlosnikFerreiraStarkman2006},\cite{Skordis:2009bf}. As in the model considered here, the longitudinal component of the vector field can source the growth of structure \cite{SkordisEtAl2006}, \cite{DodelsonLiguori2006}. We will call this field $V_{T}$. We consider a matter dominated era where $V_{T}$ is responsible for the dominant source in the Poisson and shear equations. These equations respectively then are\cite{Skordis2006}:

\begin{eqnarray}
\label{eq:tevpo}
k^{2}\Phi & \approx& -f_{s}{\cal H}k^{2}V_{T}
          -\frac{K_{B}}{2}k^{2}V_{T}' \\
k^{2}(\Psi-\Phi) & \approx& f_{s}k^{2}
                (2{\cal H}V_{T}-V_{T}') \\
f_{s}(\tau) &\equiv & \frac{(1-\bar{A}^{4})}{\bar{A}^{4}}
\end{eqnarray}

where $K_{B}$ is a positive constant of the action and $\bar{A}^{4}$ is the norm of the vector squared again (equal to unity in the fixed norm case, but in TeVeS the deviation of $|\bar{A}^{2}|$ from unity is essentially the background variation of the `scalar field' degree of freedom). Therefore:

\begin{eqnarray}
\label{eq:iswi}
\Psi+\Phi = -(f_{s}(\tau)+K_{B})V'_{T}
\end{eqnarray}

In this era the vector field equation is:

\begin{equation}
V_{T}''+b_{1}\frac{V_{T}'}{\tau}+b_{2}\frac{V_{T}}{\tau^{2}}=S[\Phi,\Psi]
\end{equation}

where

\begin{eqnarray}
b_{1} &=& 2(3-\bar{A}^{4})\\
b_{2} &=& 2(2-\bar{A}^{4})+\frac{12\bar{A}^{4}}{K_{B}}(1-\bar{A}^{4})
\end{eqnarray}

and $S[\Phi,\Psi]$ is a source term. 

For this situation to arise, there must be a growing mode in $V_{T}$ \cite{DodelsonLiguori2006}. Therefore we require that $b_{2}<0$.
The function $\bar{A}^{4}$ will be rather close to unity \cite{BourliotEtAl2006} so

\begin{equation}
b_{2}\sim 2+12 f_{s}(\tau)/K_{B}
\end{equation}

Therefore we require that $f_{s}(\tau)/K_{B} \lesssim -1/6 $ (recalling that $K_{B}$ is a positive number). By (\ref{eq:iswi}) we see that the contribution to the ISW will be proportional to $f_{s}(\tau)+K_{B}$. Meanwhile we see from (\ref{eq:tevpo}) that the comparative contribution to Poisson's equation is also independently weighted by $f_{s}$ and $K_{B}$.  If there is no time dependence in the Poisson equation  (\ref{eq:tevpo}) due to the vector field then the vector field will make no contribution to the ISW effect. If there is a time dependence in the Poisson equation, there may yet be no contribution to the ISW if $f_{s}(\tau)\sim -K_{B}$ between last scattering and now (though the overdensities of the baryon field will then not generally grow as $a$, thus contributing to the ISW effect). This would be consistent with the condition for a growing mode in $V_{T}$ but it does not guarantee that the resulting growing mode would be suitable. 

Indeed, it was found  \cite{SkordisEtAl2006} that in seeding the growth of large scale structure in TeVeS there was a significant danger of incurring unacceptably high temperature anisotropies in the CMB on large scales. Although involving a larger number of terms, the same reasoning carries over to the model considered here - i.e. parameters which allow realistic structure formation will typically lead to an unacceptable ISW effect. This is vividly illustrated in Figure \ref{dark matter spectra} which shows best fit models as compared to SDSS large scale structure data. In every case,  the corresponding temperature anisotropy displays a dramatically poor fit to the data at low $\ell$.

\subsection*{Observable 2: Amplitudes}
\label{amplitudes section}
The ratio of the observed amplitudes of the CMB anisotropy and the matter power spectrum is consistent with a growth rate proportional to $a$ (though see \cite{Bean2009}).  Any uncompensated change to this growth rate in the aether model over this period would lead to a different ratio.  The bias parameter between the galaxy distribution and the underlying density field can be used to rectify this difference, but only if the change is relatively small and unphysically large bias parameters (larger than $\sim10$) are not required.

\subsection{Summary}

We have seen that although the vector field may play a number of the roles that dark matter plays, it seemingly cannot do all at once \footnote[1]{It is interesting to compare this with similar results found in \cite{Li:2008aia} for a framework for generalizations of dark matter. It is not clear whether the model considered in this paper fits within this framework}.

The position of the acoustic peaks in the CMB temperature anisotropy should be taken as a strong indication of additional nonbaryonic nonrelativistic mattter present in the universe during matter domination. Such an effect can be achieved in this model by a rescaling of the value of Newton's constant. However, this rescaling cannot persist into the radiation era \cite{Carroll:2004ai}. Thus, the functional form would have to be approaching the rescaling solution only after one would expect matter (including cold dark matter)-radiation equality to happen.

This implies the presence of a new scale in the theory, roughly corresponding to the Hubble parameter $H_{eq}$ at this time. It seems fair to say in general that a model such as that considered here, is more likely to be a cosmologically viable candidate for dark matter if the scale $M$ in the theory is closer to $H_{eq}$ and not $H_{0}$. It is tempting to speculate
whether a theory where the scale $M$ itself is dynamical may find more success, but that will not be explored in this work.

Furthermore we have seen that even if the background is consistent with observations, the effect on the evolution of perturbations may be unacceptable, notably either through the ISW effect or comparing the respective amplitudes of the CMB anisotropy and matter power spectrum today.

\subsection{Example Problem Spectra}
It has previously been shown that the Einstein Aether can produce acceptable matter power spectra with certain parameter combinations \cite{ZlosnikFerreiraStarkman2008}.  Here we show that such combinations do not provide an acceptable fit to CMB measurements.  Despite extensive searches we have been unable to find any parameter set within the model that does fit the WMAP data well; this is entirely in line with the problems discussed above.

We use a parameter set which is consistent with BBN limits on $\Omega_B h^2$ and the HST key project measurement of $H_0$.  The standard cosmological parameters are: $\Omega_b h^2=0.0193$, $n_s=0.83$, $H_0=89.3\, \mathrm{km/s/Mpc}$, $\Omega_c h^2=0$.  The new Aether variables are $c_+=-4.72$, $c_-=-6.11$, $n_{\mathrm{ae}}=0.34$, $\Omega_{\mathrm{ae}}=0.82$, $M_{\mathrm{ae}}=111.3\,\mathrm{km/s/Mpc}$ with  $c_2$ set by requiring zero sound speed.  This parameter combination is in no sense optimal, but it does provide an illustration of all the problems that arise here.

Figure \ref{dark matter spectra} shows power spectra from our modified Boltzmann code for this parameter set.  The matter power is a realistic fit to the SDSS data (this was the criterion for our choice of parameters).  The CMB spectra shows various problems.  In the low-$\ell$ regime a large ISW effect is clearly present, destroying the fit at large scales, as described in section (\ref{ISW subsection}).  The positions of the peaks are poorly fit by the model, as expected and discussed in section (\ref{background subsection}).  Finally, in the plot we have rescaled the amplitude of the matter power spectrum by a factor $0.02$, corresponding to a galaxy bias of $0.14$ in order to reconcile the relative amplitudes of the two spectra with the data; such a scaling is unphysically small.  This corresponds to the changed growth rate described in section (\ref{amplitudes section}).  All these effects cause severe problems when attempting to simultaneously fit the CMB and large scale structure.

\begin{figure}
\begin{center}
\epsfig{file=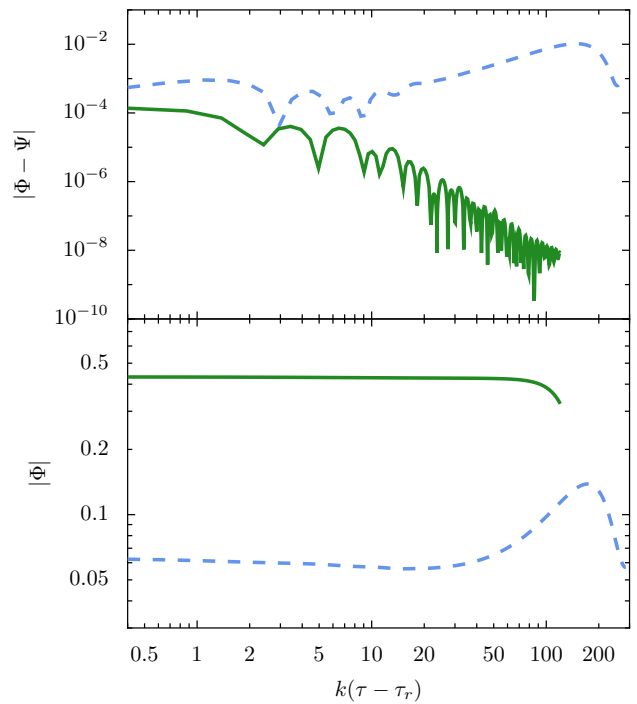,width=8.3cm,height=9.3cm}
\caption{Exotic behavior of metric potentials for $k\sim$ $10^{-2}$ $Mpc^{-1}$. The panels show the fields $|\Phi|$ and $|\Phi-\Psi|$ for a $\Lambda$CDM universe (green solid line) and GEA universe (dashed blue line) as a function of $k(\tau-\tau_r)$ where $\tau_r$ is the conformal time of recombination.
}
\label{isw1}
\end{center}
\end{figure}

\begin{figure}
\begin{center}
\epsfig{file=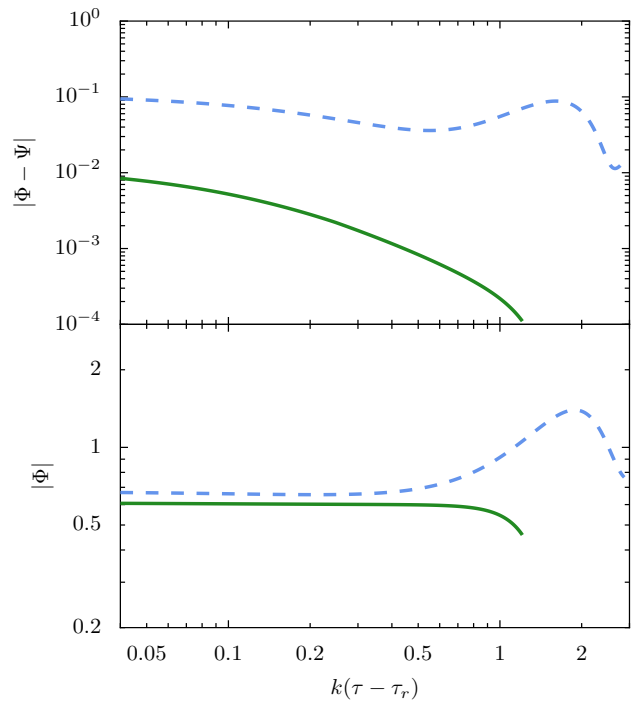,width=8.3cm,height=9.3cm}
\caption{Equivalent of Figure \ref{isw1} for k $\sim$ $10^{-4}$ $Mpc^{-1}$}.
\label{isw2}
\end{center}
\end{figure}

\begin{figure}
\begin{center}
\epsfig{file=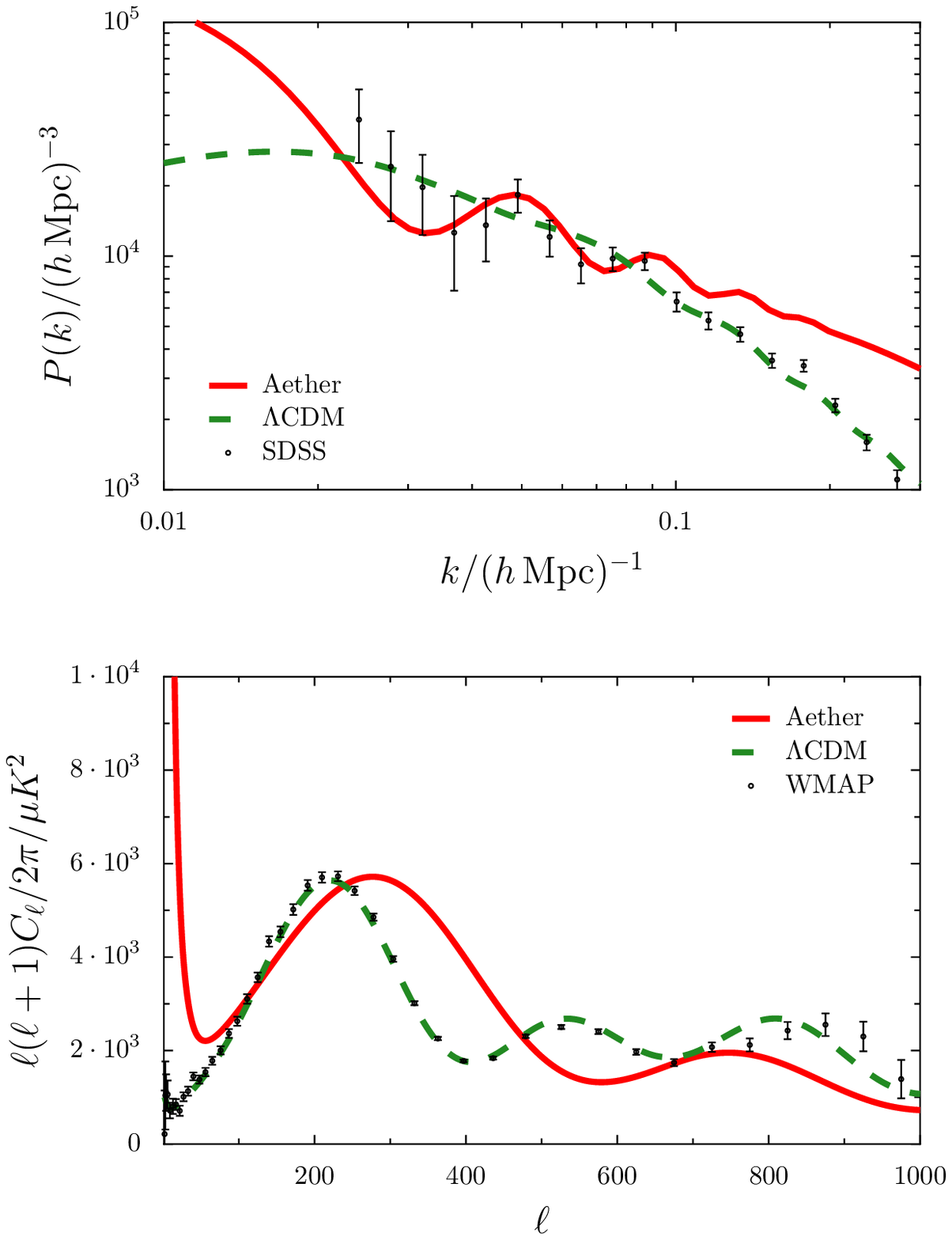,width=8.3cm,height=9.3cm}
\caption{Matter power (top) and CMB (bottom) power spectra for the $\Lambda CDM$ (dashed green) and typical GEA (solid red) models, with WMAP and SDSS constraints.
}
\label{dark matter spectra}
\end{center}
\end{figure}

Figures \ref{isw1} and \ref{isw2} illustrate the sources of the extreme ISW effects shown in Figure \ref{dark matter spectra}; the time derivatives of the metric quantities plotted create an ISW effect as shown in equation (\ref{isw equation}).  The onset of background acceleration in each case is marked by a turnover in the curves at late time. The GEA universe exhibits a dramatically increased $|\Phi-\Psi|$  and time dependence of $|\Phi|$ during the matter era as compared to the $\Lambda$CDM universe.  Although the $|\Phi-\Psi|$ has a smaller magnitude its time dependence can be significant for the total ISW effect.  Note that values of $\tau$ between the two universes do not correspond to the same physical time or redshift since the universes expand at different rates.

\section{The vector field as Dark Energy}
\label{as dark energy}
\subsection{Dark Energy Regime}
As shown in figure \ref{fig1dyn}, our vector field can produce late-time acceleration and so play the role of dark energy.  Indeed, for the form of the vector field used here, as the index $n_{\mathrm{ae}}\rightarrow0$ the theory becomes the same as a cosmological constant for both the background and the perturbations.  Since the model can fit the data well we can use our MCMC engine to find constraints on the parameters of the vector field, telling us exactly how close to the $n_{\mathrm{ae}}=0$ cosmological constant case the theory must be to fit the CMB and LSS data.

If the model can fit the data only extremely close to $n_{\mathrm{ae}}=0$ then it does not provide a compelling alternative to the cosmological constant.  If, on the other hand, there is significant flexibility in the model and no fine tuning, or if it can provide a better fit than $\Lambda$CDM, then it is somewhat more interesting.

In this section we will consider the resulting background evolution, CMB temperature anisotropy, and matter power spectrum for a universe containing the vector field, cold dark matter, the conventional matter fields, and no cosmological constant. The acceleration will arise solely from the vector field's modification to the Friedmann equation.  

\subsection{Constraints on aether dark energy from data}
\begin{figure}
\begin{center}
\epsfig{file=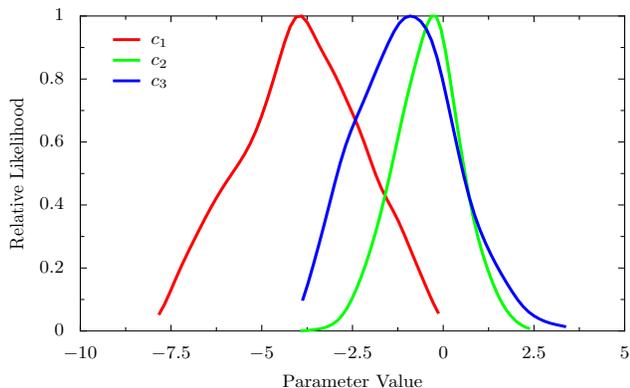,width=8.3cm,height=5.12cm}
\caption{Constraints on the three coupling terms of the theory.
}
\label{c1c2c3 constraint}
\end{center}
\end{figure}

\begin{figure}
\begin{center}
\epsfig{file=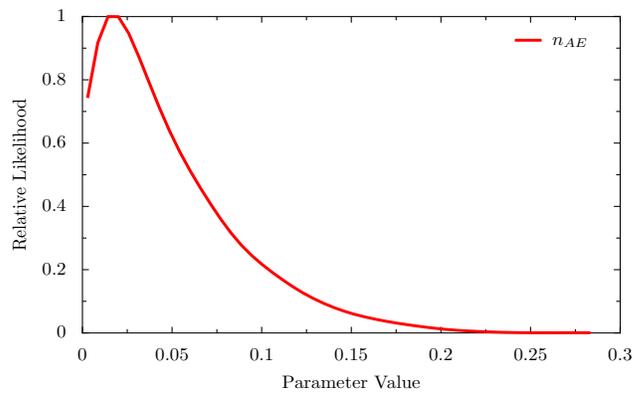,width=8.3cm,height=5.12cm}
\caption{Constraints on the kinetic term power law index parameter.
}
\label{nae constraint}
\end{center}
\end{figure}

\begin{figure}
\begin{center}
\epsfig{file=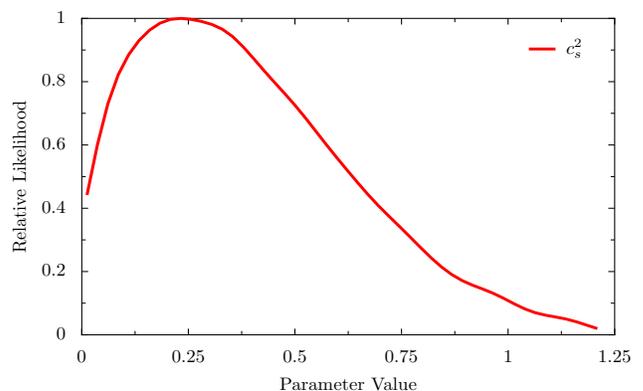,width=8.3cm,height=5.12cm}
\caption{Constraints on the vector field sound speed parameter.
}
\label{cs2 constraint}
\end{center}
\end{figure}

\begin{figure}
\begin{center}
\epsfig{file=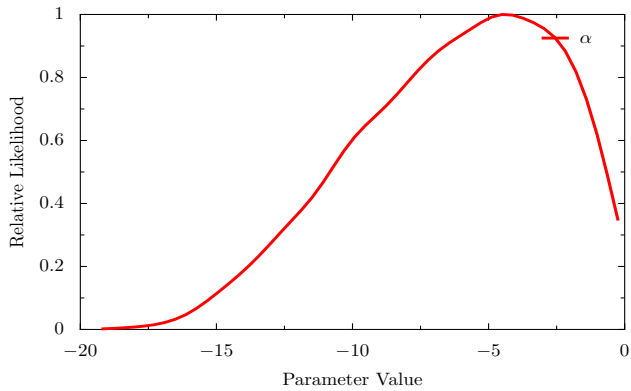,width=8.3cm,height=5.12cm}
\caption{Constraints on the parameter $\alpha=c_1+3c_2+c_3$
}
\label{alpha constraint}
\end{center}
\end{figure}

Our MCMC generated the constraints on the vector field parameters shown in figures \ref{c1c2c3 constraint} to \ref{alpha constraint}; these curves are the smoothed histograms from our combined Markov chains.

The most important trend evident in these results is closeness of $n_{\mathrm{ae}}$ to the $\Lambda$ value of zero.  We find the best fit value $n_{\mathrm{ae}}=4.2\cdot10^{-3}$, with a 95\% upper limit $n_{\mathrm{ae}}<0.126$.

The best fit value in the MCMC run is very close to the $\Lambda$CDM likelihood of the same data, at the cost of six extra parameters, meaning that it is unlikely to be favored by any model comparison exercise.  It does, however, demonstrate the validity of modified gravity-related dark energy candidates.

Having obtained these constraints we can determine their origin.  There are two ways in which the vector field must behave like $\Lambda$ to provide a good fit.  The first is that the late-time acceleration should be close to that given by $\Lambda$.  The second is that any perturbations in the field (which are not present in $\Lambda$) should not affect the observable spectra.

\subsection{Constraint origins - acceleration rates}

The consistency of the acceleration of the universe with the cosmological constant equation of state $w=-1$ is being measured with increasing precision in supernova and baryon acoustic oscillation experiments (which are beyond the scope of this paper).  Here, they will be constrained by the late-time ISW effect induced by dark energy, and by the perturbation growth rate.

We can assess how closely vector-induced acceleration mimics $\Lambda$-driven expansion at late times with the equation of state $w_{\mathrm{ae}}$ of the vector field in the background:
\begin{equation}
w_{\mathrm{ae}} = -1-\frac{1}{3H^{2}}\frac{\frac{d^{2}}{dtdH}F}{\frac{d}{dH}
\left(\frac{F}{H}\right)}.
\end{equation}

For the monomial form of $F(K)$ we have that:

\begin{eqnarray}
\label{waez}
w_{\mathrm{ae}} = -1 -\frac{2n}{3(2n-1)}\frac{\dot{H}}{H^{2}}
\end{eqnarray}

\begin{figure}
\begin{center}
\epsfig{file=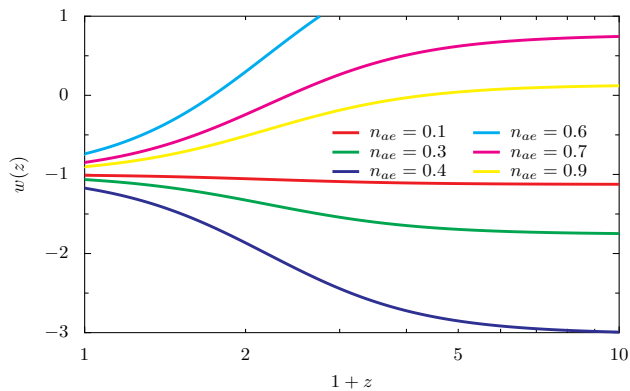,width=8.3cm,height=5.12cm}
\caption{The vector field's equation of state as a function of redshift $z$, for various values of the kinetic term index $n_{\mathrm{ae}}.$
}
\label{wvsz}
\end{center}
\end{figure}

Thus the equation of state will generically deviate from $-1$ whenever $n \neq 0$ and so the acceleration for these values will not be degenerate with a cosmological constant. We see immediately from equation (\ref{waez}) that  $w_{\mathrm{ae}}(\tau) < -1$ for $ 0 < n < 1/2$ and $w_{\mathrm{ae}}(\tau)> -1 $ for $n> 1/2$. This is clearly visible in Figure \ref{wvsz}.

\subsection{Constraint origins - perturbation evolution}

Even if the background expansion is rather close to the the $\Lambda$CDM model, the evolution of perturbations need not be.  
This is most easily illustrated by considering the Poisson equation on large scales (see  \cite{ZlosnikFerreiraStarkman2007} for a derivation).  On these large scales there is a time-dependent re-scaling of the the metric perturbation $\Phi$, which we can cast as a modification of the effective gravitational constant $G$:

\begin{eqnarray}
k^{2}\Phi &=& -4\pi G^{(1)}_{\mathrm{eff}} a^{2} \sum_{i}\bar{\rho}_{i}\delta_{i} \\
G^{(1)}_{\mathrm{eff}}  &\equiv& \frac{G}{1+\frac{c_{1}}{2}F_{K}}
\end{eqnarray}

where we have assumed that terms proportional to the velocity divergence are ignorable and provisionally considered the effect of the field $\xi$ to be subdominant.

The Friedmann equation may be used to cast the above equation in a more familiar form by eliminating he background $\rho_{i}$ in favour of background expansion rate of the universe and the time-dependent fractional energy density $\Omega_{i}(\tau)\equiv 8\pi G \rho(\tau)/(3H(\tau)^{2})$. This yields:

\begin{eqnarray}
k^{2}\Phi &=& -\frac{3}{2}{\cal H}^{2}\frac{G^{(1)}_{\mathrm{eff}} }{G^{(0)}_{\mathrm{eff}} } \sum_{i}\Omega_{i} (\tau)\delta_{i} 
\end{eqnarray}

where

\begin{eqnarray}
G^{(0)}_{\mathrm{eff}}  &\equiv& \frac{G}{1-\alpha K^{\frac{1}{2}}\frac{d}{dK}\left(\frac{F}{K^{\frac{1}{2}}}\right)}
\end{eqnarray}

\begin{figure}
\begin{center}
\epsfig{file=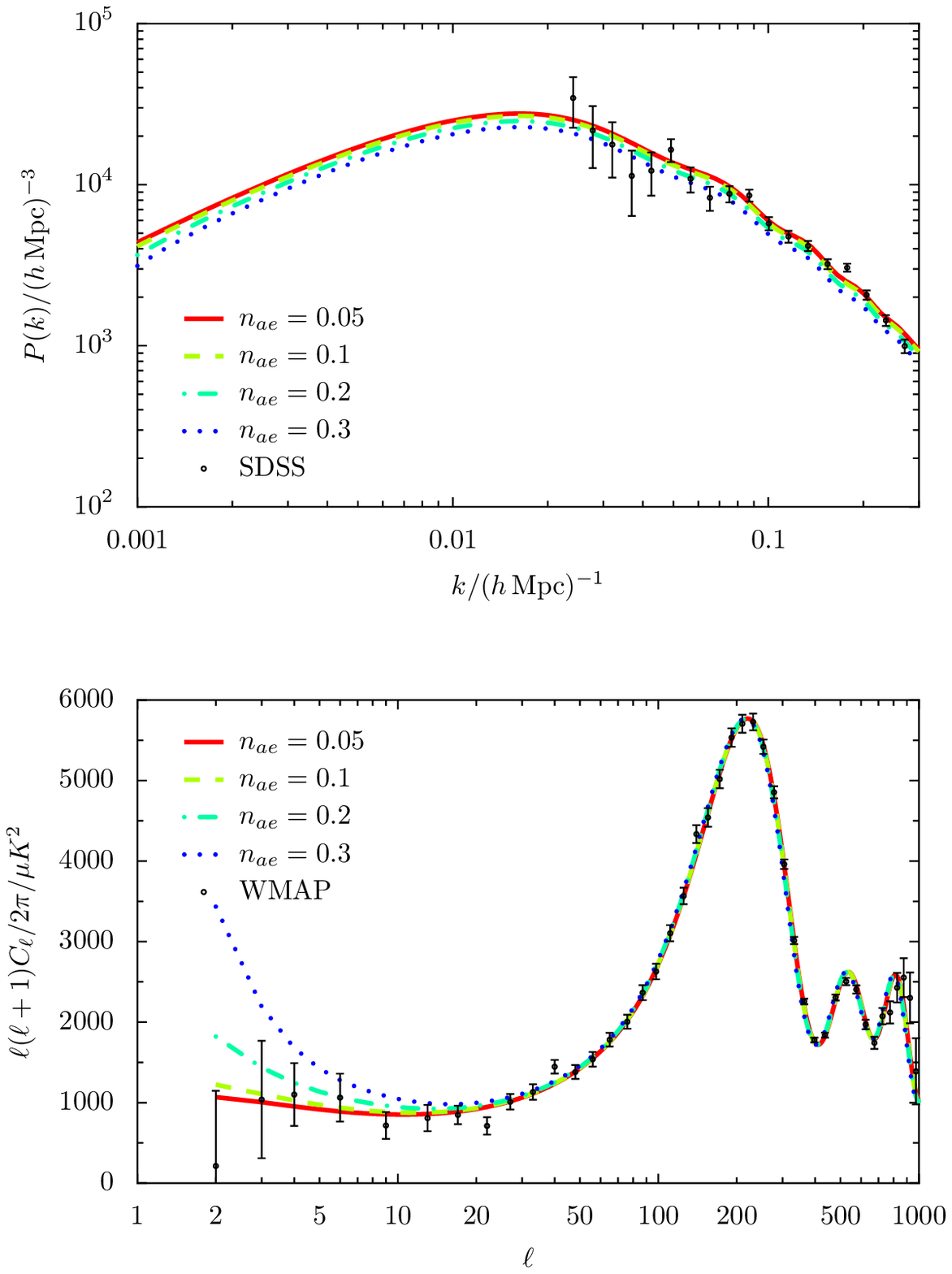,width=8.3cm,height=9.3cm}
\caption{Matter power (top) and CMB (bottom) power spectra for the various GEA dark energy models, with WMAP and SDSS constraints. The power-law function's exponent $n$ is varied}.
\label{nae}
\end{center}
\end{figure}

The $n=0$ $\Lambda$CDM Poisson equation may be cast in this form by taking  $G^{(1)}_{\mathrm{eff}}= G$ and $G^{(0)}_{\mathrm{eff}}=G/(1-\Lambda/(3H^{2}))$.  
For the case where $n$ differs from $0$,  the function $G^{(1)}_{\mathrm{eff}}$ will generically possess a time dependence during the background evolution. Therefore the link between the time evolution of the functions $G^{(0)}_{\mathrm{eff}}$, $\delta_{i}$ and $\Phi$ will differ from the case where acceleration is caused by a cosmological constant. We may thus expect the ISW effect to be of a non-standard form. This is vividly illustrated in Figure \ref{nae} where it can be seen that for a given set of $(c_{i},\gamma,M)$, variation of $n$ results in a considerable variation in the large scale CMB temperature anisotropy. Also evident is a variation in the matter power spectrum amplitude with $n$, evident  on all scales.

\begin{figure}
\begin{center}
\epsfig{file=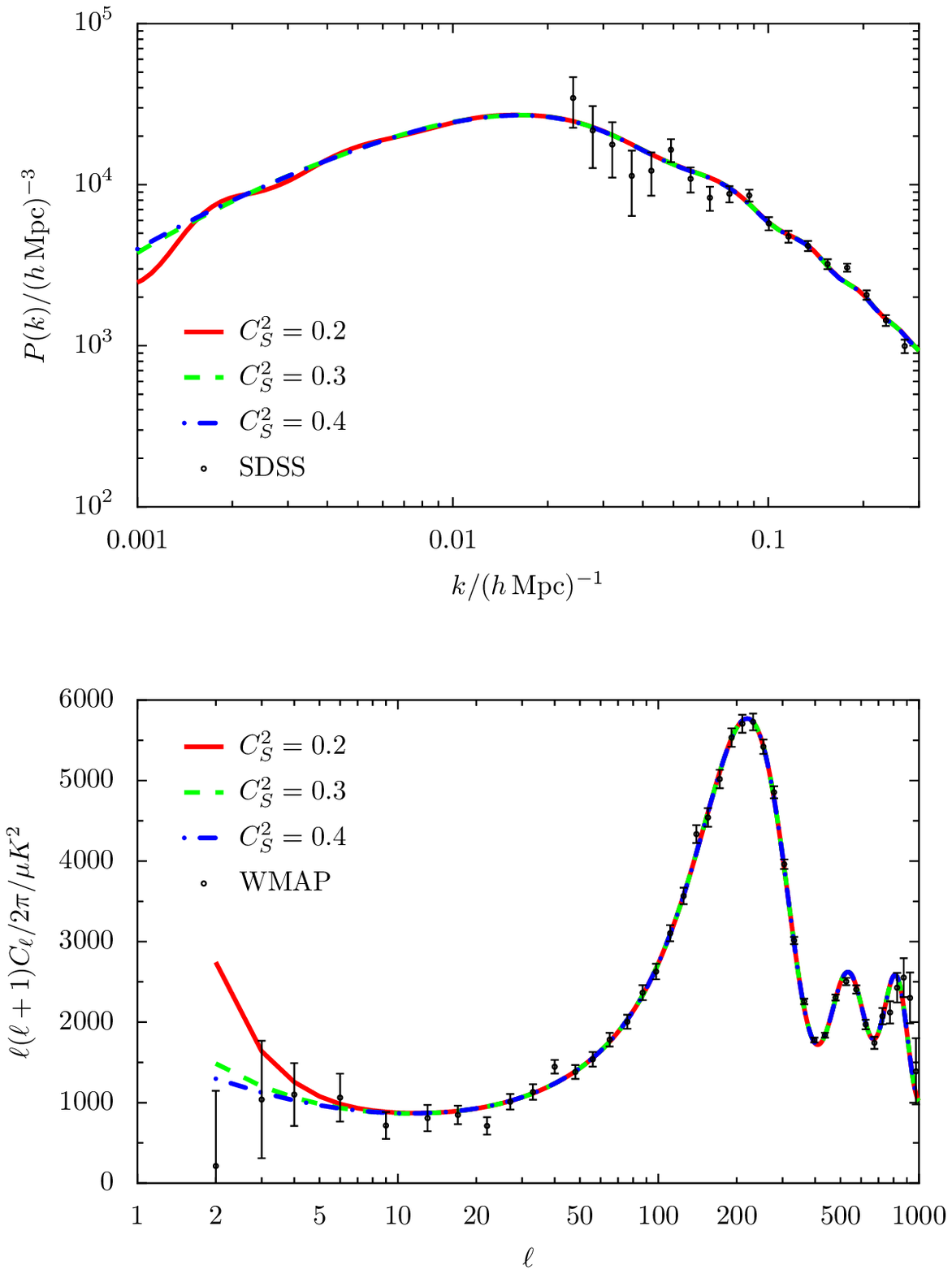,width=8.3cm,height=9.3cm}
\caption{Matter power (top) and CMB (bottom) power spectra for the various GEA dark energy models, with WMAP and SDSS constraints. The squared speed of sound of vector field perturbations is varied.
}
\label{cs2}
\end{center}
\end{figure}

Variation of parameters other than $n$ could also have a significant impact on the success of the models. Given the results of the previous section, it seems unlikely that any influence of the field $\xi$ would tend to improve the models. This indeed seems to be the case. Figure \ref{cs2} depicts various models where the function is varied $C_{S}^{2}$ for fixed values of the other parameters. In particular, $n$ takes the value $0.1$. The function $C_{S}^{2}$ is ultimately a measure of the ability of the field $\xi$ to sustain any homogeneous growing behavior for $(k\tau)>1$; higher values will tend to limit the effect of the vector field to larger and larger scales. The sets of parameters were chosen so such a growing solution indeed existed on superhorizon scales. The figure indicates that a growing $\xi$ field will indeed have deleterious effects on scales where it is not suppressed. There is a significant ISW effect evident for the red curve CMB; the corresponding model must be considered as being at the edge of acceptability. The corresponding 
large scale matter power spectrum exhibits exotic oscillations, entirely unrelated to the baryon-acoustic oscillations which occur on other scales. Their presence is in this model is thus reflective of dynamics in the dark energy sector.

\section{Conclusions}
\label{conclusions}

\subsection{Being Dark Matter is hard}
Generalized Einstein-Aether can, with different parameter choices, resemble dark matter in some important ways but never all of them at once.  Specifically, the new degrees of freedom introduced by the model may conspire to identically replicate one or more but not all of the following properties of cold dark matter.

\subsubsection{Background dynamics}
To accomplish identical background dynamics to cold dark matter, one must introduce considerable fine tuning into the function $F$ of the theory. Specifically the function must change form either side of (dark) matter-radiation equality.  The parameter we tune to make this happen is in the action itself, unlike the usual case where we simply alter the abundance $\Omega_c$.  Changes to fit cosmological observations can therefore have a larger impact on the small scale behavior of gravity.

\subsubsection{The speed of sound}
If the speed of sound is too high then structure cannot form on small enough scales.  In this model we may reduce the sound speed to be close to zero, at the cost of one of our parameters.  When designing new gravity theories this is perhaps the easiest structure formation constraint to investigate, and it should be examined to see if it conflicts with other constraints needed to make the theory useful - for an example, see \cite{Seifert:2007fr}. 

\subsubsection{Growth rate of `overdensity'}
Theories of modified gravity designed to replace dark matter must necessarily have growing modes of fluctuation in at least one of the new degrees of freedom they introduce, in order to sufficiently source gravitational collapse and structure formation on scales within their own sound horizon.  
There remains some flexibility in the perturbation growth rate, since the bias on the galaxy power spectrum measurements is a free factor.  As measurements of weak lensing (which samples gravitation directly) and semi-analytic models (which predict bias) improve this freedom will be reduced.

\subsubsection{Absence of anisotropic stress and contribution the cosmological Poisson equation}
A sufficiently small anisotropic stress associated with the vector field may be implemented by fine tuning the parameter $c_{13}$ to be very small (see equation (\ref{shear})).  However, as was discussed in Subsection \ref{ISW subsection}, even this will tend to come at the expense of other desired behavior of the field.
An appreciable time variation of the anisotropic stress over the time from last scattering to today can result in a very poor fit to the 
low-$\ell$ CMB $C_\ell$.  This problem is likely to be common in theories of modified gravity.

As we have seen, it is a combination of time variation of the anisotropic stress and time variation of the field $\Phi$ via the vector field's effect on the cosmological Poisson equation that contribute to the ISW effect. Though both effects are absent in the cold dark matter case, one may imagine both effects being present in the vector field model but being of equal and opposite sign. As with the case of the isolated anisotropic shear contribution, it seems this is not possible whilst maintaining the other constraints like the existence of a growing mode.

\subsubsection{Effective minimal coupling to the gravitational field}
Even if the vector field growing mode gives an appropriate (dark matter-like) contribution to the Poisson equation, the link between the overdensity and the corresponding $\Phi$ can differ from the CDM case.  This difference can come from curvature terms in the vector field stress energy tensor, and its main consequence is a time-dependent rescaling of Newton's constant $G$.  Thus $\Phi$ may gain a time dependence during the matter era even there is a completely standard dark matter contribution to the Poisson equation.  The converse may also be possible:  the time dependence of an incorrect Poisson contribution could be counteracted by a time dependence of the effective $G$.

\subsection{Being Dark Energy is easy}

As a model for dark energy the Generalized Einstein-Aether theory is more successful: we have obtained constraints on its parameters and found that it generates spectra that fit the data across a reasonable range of its parameter space.  It is clear that modified gravity approaches to explaining late-time acceleration are viable and can provide motivated explanations for dark energy (though this theory retains the co-incidence problem in the guise of the parameter $M$).

\subsubsection{Closeness to $\Lambda$}
The most interesting constraint on this branch of the theory is on the parameter $n_{\mathrm{ae}}$ and is shown in figure \ref{nae constraint}.  In some sense this parameter describes how closely the theory mimics $\Lambda$ (which has $n_{\mathrm{ae}}=0$).  The fact that this parameter is rather free, $n_{\mathrm{ae}}<0.126$, (95\% CL), is consistent with the fact that a wide variety of other theories can also explain dark energy: present structure formation data is not very informative about the nature of dark energy, and deeper require expansion probes like baryon acoustic oscillation and supernovae.

\subsubsection{Sound speed}
The other notable constraint on Einstein-Aether dark energy, which may extend to other modified gravity approaches, is the limits on the sound speed, illustrated in figure \ref{cs2 constraint}.  As shown in figure \ref{cs2}, an incorrect sound speed can lead to large scale oscillations by modifying the other parameters of the theory and permitting a growing mode excitation at late time.

\subsubsection{Other constraints}
The other constraints on the theory (which are easily fulfilled by choosing the $c$ parameters) come from ensuring that no growing mode can disrupt the power spectra, that the acceleration is close to the $\Lambda$CDM value, and that the value of effective $G$ remains positive at all times.

\subsection{Future issues for modified gravity and structure formation}
Because $\Lambda$CDM is such a good fit to current cosmological data, modified gravity will never be favoured in a model comparison exercise using only current data about linear structure. It is only in combination with physics on galactic and smaller scales that it can be be persuasive.  This work highlights a few issues for future model-building in this vein.

The Generalized Einstein Aether model is a member of a class of models in which the scale $M \sim H_0$ associated with Dark Energy is visible to dark matter.  Many of the issues raised here will be relevant to any such models which try to use a dark matter scale consistent with small-scale modifications to gravity.

A combination of probes sensitive only to the background (like Type 1A supernovae) and to the behavior of cosmological perturbations is needed to fully constrain these theories.  For example, a value $n_{\mathrm{ae}}=0.3$ has $w(z)\sim 1$ at low redshift but is ruled out by our constraints.  Similarly there are models with $n_{\mathrm{ae}}\sim 0.75$ that provide reasonable spectra, but they are ruled out by $w(z)$ constraints.

There are, of course, a number of extensions to the theory and features of it that could be change; we could, for example, allow $M$ to vary dynamically, or add more terms to the kinetic component $F$ in equation (\ref{kinetic equation}).  There are also myriad possibilities in more changing more general aspects of the primordial conditions or cosmological parameters: what happens if we add add tensors?  Can we include an isocurvature mode?  Would massive neutrinos help?  Or curvature? This leads us to a key caveat that applies to this and all similar work constraining new physics with linear structure:  a simple constraint from the data alone is worthless, since any of the numerous other parameters we could change might conspire to counteract whatever problem it solves.  We need a physical explanation of a constraint's origin to understand whether it is robust to the cosmologist's tinkering.

{\it Acknowledgments}:
We thank Constantinos Skordis and David Jacobs for useful discussions.  GDS was supported by a grant to the CWRU particle/astrophysics theory group from the US DOE.    JZ is supported by an STFC rolling grant.  TGZ is supported by Perimeter Institute for Theoretical Physics. Research at Perimeter Institute is supported by the Government of Canada through Industry Canada and by the Province of Ontario through the Ministry of Research \& Innovation.
\bibliography{references}

\appendix 
\section{The MOND regime F(K) cannot yield late-time acceleration}
\label{mond regime bad}
The MOND branch of $F$ applies in the regime $0\le K\ll1$; outside this range MOND does not prescribe its form.  Inside that range the MOND value is:
\begin{eqnarray}
c_{1}F &=& -2K+ \frac{2}{3}K^{\frac{3}{2}} \\
\nonumber  ( 0 \leq &K&  \ll 1 ) \label{MOND F}
\end{eqnarray}
The weak field limit of the Einstein Aether theory is:
\begin{equation}
K^{\mathrm{ae}}_{\mathrm{WF}} = -c_{1}\frac{(\nabla\phi)^{2}}{M^{2}},
\end{equation}
where $\phi$ is the conformal Newtonian potential.  The weak field MOND limit is:
\begin{equation}
K^{\mathrm{MOND}}_{\mathrm{WF}} = \frac{(\nabla\phi)^{2}}{a_0^{2}}.
\end{equation}
Clearly we can equate these by setting M$^2=-c_1 a_0^2$.  Using this relation with equation (\ref{KFRW equation}) gives us this value for the present-day cosmological $K$:
\begin{equation}
K_{\mathrm{FRW}}(t_0)= -3\frac{\alpha}{c_{1}}\frac{H_{0}^{2}}{a_{0}^{2}}
\end{equation}

There are now two cases: $\alpha/c_1<0$ and $\alpha/c_1>0$.  In either case we need the measured values of $H_0$ and $a_0$; they are suggestively similar:
\begin{eqnarray}
H_0 &\equiv& \kappa a_0 \nonumber \\
\kappa &\approx& 6
\end{eqnarray}
where 

If $\alpha/c_1<0$ then $K>0$ and we are directly in the MOND regime.  Then we obtain the modified Friedman equation:
\begin{equation}
H^{2}\left(1+\frac{\alpha}{c_{1}}+\frac{3}{2}\kappa^{2}\left(\frac{H}{H_{0}}\right)\left(\frac{-3\alpha }{c_{1}}\right)^{\frac{3}{2}}\right)=8\pi G \rho
\end{equation}
A self-accelerating solution to this is only possible if the quantity in brackets is positive definite, so that $H\rightarrow$ \emph{const} as $\rho \rightarrow 0$.  This could only happen if $\alpha/c1<-1$, but that would violate our requirements that $K\ll1$.  

In the other case $\alpha/c_1>0$ we must extrapolate the MOND form of $F$ in equation (\ref{MOND F}) to $K<0$.  To make the extension continuous across the $K=0$, we should set $F(-K)=-F(K)$, so that the MOND form becomes:
\begin{eqnarray}
c_{1}F &=& -2K- \frac{2}{3}(-K)^{\frac{3}{2}} \\
\nonumber  ( 0 \leq &-K&  \ll 1 ) \label{MOND F negative}
\end{eqnarray}

The Friedman equation then becomes:
\begin{equation}
H^{2}\left(1+\frac{\alpha}{c_{1}}-\frac{3}{2}\kappa^{2}\left(\frac{H}{H_{0}}\right)\left(\frac{3\alpha }{c_{1}}\right)^{\frac{3}{2}}\right)=8\pi G \rho
\end{equation}
which does have an accelerating solution.  Unfortunately, solving this equation at the present day for reasonable values of $\Omega_m$ shows that it requires $K\sim 2$, which again violates our requirement that $|K|\ll1$.   Having exhausted our other options we are forced to require $K\ge1$, outside the true MOND regime.  This is consistent with another separate analysis of the solutions of (\ref{fried}) \cite{Cardone:2009hs}. It seems likely that if the vector field is responsible for the late time acceleration then it is a result of behavior of the function
away from the MONDian limit.  There is still, though a role for the near numerical coincidence of $a_{0}$ and $H_{0}$ - it lessens the fine tuning of the other parameters in $F$ in order for the acceleration to happen at suitably late times.

\section{Vector And Tensor Modes}
\label{vt}

In this appendix, we provide the perturbed equations of motion for the vector and tensor modes of the various fields appearing in these models, namely the metric, the vector field and the Lagrange multiplier. The latter, being scalar, only have a spin$-0$ mode. The other two fields, $g_{ij}$ and $A^i$ are perturbed as
\bea
ds^2 & = & -a^2d\tau^2 +a^2 B_i d\tau dx^i+a^2(\gamma_{ij}+h_{ij})dx^i dx^j \nonumber\\
h_{ij} &=& 2\partial_{(i} E_{j)}+2E_{ij}\nonumber\\
A^{\mu} & = & \left(\frac{1}{a},\frac{A^i}{a}\right),\nonumber\\
A_{\mu} & = & \left(-a,aV_i\right),
\eea
where the different fields introduced satisfy
\be
\partial_i B^i=0=\partial^j E_{ij}=\partial_i E^i=E^i_i \text{ and } \partial_i A^i=0.
\ee
We also introduce the useful quantity $V^i=A^i+B^i$ and remind that all the latin indices on the perturbed fields are raised and lowered thanks to the Kronecker flat metric $\delta_{ij}$.\\

In the following we use the unperturbed results
\bea
K & = & 3M^{-2}\mathcal{H}^2\alpha,\nonumber\\
J^0_{\ 0}=6c_2\mathcal{H}& , & J^i_{\ j}=2\mathcal{H}\alpha\delta^i_j,\nonumber\\
I^{\sigma}_{00}=6c_2\mathcal{H}\delta^{\sigma}_0 & , & I^{\sigma}_{0i}=0,\ I^{\sigma}_{ij}=-2\mathcal{H}\alpha\delta^{\sigma}_0\delta_{ij}.
\eea
The Einstein Equation (\ref{Einsteineq1}) without matter introduces the stress tensor
\bea
\tilde{T}_{\alpha \beta} &=&\frac{1}{2}\nabla_{\sigma}\left[I^{\sigma}_{\alpha \beta}\right]+\hat{T}_{\alpha\beta} \nonumber \\
\nonumber  \hat{T}_{\alpha\beta}&=&-\cf Y_{(\alpha\beta)}+\frac{1}{2}M^2 g_{\alpha\beta}\mathcal{F}+\lambda A_{\alpha}A_{\beta},
\eea
where
\be
I^{\sigma}_{\alpha \beta}=\cf \left[J_{(\alpha}^{\ \ \sigma}A_{\beta)}-J^{\sigma}_{\ (\alpha}A_{\beta)} -J_{(\alpha\beta)}A^{\sigma}\right].
\ee
As intermediate results we have
\be
\delta K^{\alpha\beta}_{\ \ \ \gamma\delta}=c_1\left(\delta g^{\alpha \beta}g_{\gamma\delta}+g^{\alpha\beta}\delta g_{\gamma\delta}\right),
\ee
\be
\delta \left(\nabla_{\alpha} A^{\gamma}\right)=\delta^{\gamma}_i\partial_{\alpha}\left[ \frac{A^i}{a}\right]+\Gamma^{\gamma}_{\alpha k} \frac{A^k}{a}+\frac{\delta \Gamma^{\gamma}_{\alpha0}}{a},
\ee
\be
\delta Y_{0i}=\delta Y_{i0}=c_1\left(\mathcal{H}V_i +V_i'\right),
\ee
\bea
\delta \hat{T}_{ij} &=&\frac{1}{2}M^2 F a^2h_{ij}, \nonumber \\
\delta \hat{T}_{0i}&=&\frac{1}{2}a^2 B_i M^2 F-\lambda a^2 V_i -\cf \delta Y_{(0i)},
\eea
\bea
\delta J^0_{\ i} &=& \frac{2\mathcal{H}c_3V_i}{a}-\frac{2c_1V_i'}{a}, \nonumber \\
\delta J^{\ i}_0 &=& -\delta J^0_{\ i} +\frac{2\mathcal{H}}{a}(c_1+c_3)B^i,
\eea
\bea
\delta J^{\ 0}_i &=& -\delta J^j_{\ 0}\delta_{ij} +\frac{2\mathcal{H}}{a}(c_1+c_3)B_i, \\
\delta J^i_{\ 0} &=& 2\frac{c_{13} B^i}{a}\mathcal{H}-2\frac{c_1\mathcal{H}}{a}V^i +\frac{2c_3V^{i'}}{a};
\eea
and
\bea
\delta J^i_{\ j} &=& 2c_1\left[\frac{\partial^i A_j}{a}+\frac{h^{i'}_j}{2a}+\frac{\partial^{[i}B_{j]}}{a}\right], \nonumber \\
 &&+2c_3\left[\frac{\partial_j A^i}{a}+\frac{h^{i'}_j}{2a}+\frac{\partial_{[j}B^{i]}}{a}\right].
\eea
We then obtain,
\bea
\delta I^{\sigma}_{0i} & = & \cf\left[2\mathcal{H}c_3 V_i -2c_1 V_i' -2\mathcal{H}\alpha B_i\right]\delta^{\sigma}_0 \\
\nonumber && +2\cf (c_1-c_3)\partial^{[k}V_{i]}
\eea
with $\alpha=c_1+3c_2+c_3$, and
\bea
\delta I^i_{00}&=&\cf[-2c_1V^{i'}-2c_1\mathcal{H}V^i+2\mathcal{H}c_3V^i\nonumber\\
&&+2c_3V^{i'}+6c_2\mathcal{H}A^i].
\eea
We also get
\be
\delta I^{\sigma}_{ij}=-\delta^{\sigma}_0 \cf \delta J_{(ij)}.
\ee
To obtain the perturbed Einstein equation, we then plug these results into the relations
\bea
\delta \nabla_{\sigma}I^{\sigma}_{0i} & = & \partial_{0}\left(\delta I^0_{0i}\right)+\partial_{k}\left(\delta I^k_{0i}\right) +2\mathcal{H}\delta I^0_{0i} \nonumber \\
&& -(B^{j'}+\mathcal{H}B^j)I^0_{ji}-6\mathcal{H}^2c_2B_i\cf-\Gamma^0_{ji}\delta I^j_{00},\nonumber\\
\delta \nabla_{\sigma}I^{\sigma}_{ij} & = & \partial_{0}\left(\delta I^0_{ij}\right)+2\mathcal{H}\delta I^0_{ij}-I^0_{kj}\delta \Gamma^k_{0i}-\mathcal{H}\delta_{ki}\delta I^k_{0j}\nonumber \\
\nonumber && -\delta\Gamma^k_{0j} I^0_{ik}-\mathcal{H}\delta_{kj}\delta I^k_{i0}.
\eea
We decompose these equations into Fourier components:
\bea
X^i(t,\overrightarrow{x}) & = & \sum_{\overrightarrow{k}} \sum_{m=0,1} X(t,\overrightarrow{k})Y^{i(\pm m)}_{\overrightarrow{k}},\nonumber\\
T^{ij}(t,\overrightarrow{x}) & = & \sum_{\overrightarrow{k}} \sum_{m=0,1,2} T(t,\overrightarrow{k})Y^{ij(\pm m)}_{\overrightarrow{k}}.
\eea
where the orthonormal modes
\be
Y^{(0)},\ Y^{(\pm 1)},\ Y^{(\pm 2)}
\ee
are eigenmodes of the Laplace-Beltrami operator: $\Delta Y_{\mathbf{I}}^{(m)}=-k^2 Y_{\mathbf{I}}^{(m)}$, $\mathbf{I}$ being an arbitrary set of Lorentz indices. For more information on these functions, see \cite{Lim:2004js}. The perturbed vector field equation can then be written
\bea \label{eqvec}
2\lambda a V & = & \ccf K'\left(2\mathcal{H}c_3V-2c_1V'\right)-2c_1\cf V''\nonumber \\
&&-4\cf c_1\mathcal{H}V'+V\left(2\frac{a''}{a}c_3 \cf+\mathcal{H}^2(2c_3+2c_1)\cf\right)\nonumber\\
& & -\cf \left(2c_1k^2 A+(c_1-c_3)k^2 B\right) +3\cf c_2 k^2 E'.
\eea
The spin$-2$ part of the $ij$ Einstein equations gives
\be \label{spin2ij}
E^{''}+2\mathcal{H}E^{'}+\frac{k^2 E}{1+\cf(c_1+c_3)}+2\ccf K'(c_1+c_3)E^{'}=0,
\ee
while the spin$-1$ part of the ${}^0_i$ Einstein equations gives the same equation as (\ref{eqvec}) and the spin$-1$ component of the $ij$ Einstein equation is, in space conventions,
\bea \label{spin1ij}
0 & = & \partial_{(i} E_{j)}^{''}+\frac{2(1-2\alpha)\mathcal{H}+2c_{13}\frac{1}{a^{2}}(a^{2}F_{K})' +4\alpha\mathcal{H}\cf }{1+2\cf c_{13}} \partial_{(i} E_{j)}^{'} \nonumber \\
&&-\frac{\partial_{(i}B_{j)}^{'}-c_{13}\cf\partial_{(i}A_{j)}^{'}}{1+2\cf c_{13}}\nonumber\\
& & -\frac{2(2\mathcal{H}'+\mathcal{H}^2)+M^2\mathcal{F}a^2+4\alpha({\cal H}F_{K})'+8\mathcal{H}^2\alpha\mathcal{F}}{1+2\cf c_{13}}\partial_{(i}E_{j)}\nonumber \\
&& -\frac{2\mathcal{H}\partial_{(i}B_{j)}}{1+2\cf c_{13}}\nonumber\\
& & +\frac{c_{13}\ccf K'+2c_{13}\mathcal{H}\cf}{1+2\cf c_{13}}\partial_{(i}A_{j)}.
\eea
In order to compare with the results of \cite{Lim:2004js} let us consider the gauge $E_j=0$. The spin$-1$ part of the $ij$ Einstein equations (\ref{spin1ij}) implies
\be \label{constraint}
B_i= c_{13}\cf A_i.
\ee
After that we can rewrite (\ref{eqvec}) as
\bea \label{eqvecF}
A^{''} & = & - \left(2\mathcal{H}+\frac{\ccf}{c_1\cf}K^{'}\right)A^{'}-\frac{\cf}{2c_1} \frac{2c_1+\cf(c_1^2-c_3^2)}{1+\cf c_{13}}k^{2} A \nonumber \\
&  & + \left[\mathcal{H}\frac{c_3\ccf}{c_1\cf}K^{'}+\frac{a''}{a}\frac{c_3}{c_1}+\mathcal{H}^2\frac{c_{13}}{c_1}-\frac{\lambda a}{\cf c_1}\right]A\nonumber
\eea
and obtaining the exact expression of $\lambda$ in the case $F(K)=K$,
\be
\lambda=3c_{13}\frac{\mathcal{H}^2}{a}-3c_2\frac{a''}{a^2}+6c_2\frac{\mathcal{H}^2}{a},
\ee
implies in this latter case
\bea
c_1\left(2\mathcal{H}A'+A''\right) & = & -\frac{1}{2}\left[c_1-c_3+\frac{c_{13}}{1+c_{13}}\right]k^2 A-\nonumber\\
& & \left[2\alpha\mathcal{H}^2-\alpha\frac{a''}{a}+c_1\frac{a''}{a}\right]A.
\eea

It is interesting to find that in the limit where $\mathcal{F}(K)=K$ we recover results from \cite{Lim:2004js} for (\ref{spin2ij},\ref{eqvecF}) and (\ref{constraint}).

A quick look at (\ref{spin2ij}) and (\ref{eqvecF}) implies the existence of two different speeds
\begin{eqnarray}
C_T^2&\equiv&\frac{1}{1+\cf c_{13}}, \nonumber \\
C_V^2&\equiv&\frac{\cf}{2c_1}\frac{2c_1+\cf(c_1^2-c_3^2)}{1+\cf c_{13}},
\end{eqnarray}
being the sound speeds respectively of the tensor perturbation $E_{ij}$ and the vector perturbation $A_i$. These speeds must be positive, so that it implies constraints on the parameters of the theory and the function $F(K)$. In particular, $\cf\neq \text{const.}$ will imply non-trivial constraints between the fields of the theory and the parameters $c_i$:
\be\left\{
\begin{array}{ccc}
1+\cf c_{13} & > & 0\\
\cf\left[1+\cf \frac{c_1^2-c_3^2}{2c_1}\right] & > & 0
\end{array}\right.
\ee

\end{document}